\documentclass[
reprint,
superscriptaddress,
showpacs,preprintnumbers,
amsmath,amssymb, 
aps,
notitlepage,
]{revtex4-1}

\usepackage{graphicx}% Include figure files
\usepackage{dcolumn}% Align table columns on decimal point
\usepackage{bm}% bold math
\usepackage{float}% float
\usepackage{hyperref}% add hypertext capabilities
\newcommand{\ecm}{E_{\mathrm{CM}}}
\newcommand{\x}{X(3872)}
\newcommand{\mevcc}{\mathrm{MeV}/c^2}
\newcommand{\gevcc}{\mathrm{GeV}/c^2}
\newcommand{\chidoffour}{\chi_{\mathrm{4C}}^2/\mathrm{dof}}
\newcommand{\chidoffive}{\chi_{\mathrm{5C}}^2/\mathrm{dof}}

\newcommand{\FINALBULZERO}{19}

\newcommand{\FINALBONE}{0.88^{+0.33}_{-0.27}\pm0.10}

\newcommand{\FINALBSIGONE}{5.2}

\newcommand{\FINALBULTWO}{1.1}

\newcommand{\FINALBSIGCHICJ}{4.8}
\newcommand{\FINALNCHICJ}{16.9^{+5.2}_{-4.5}}
\newcommand{\TOTALNJ}{15.1^{+4.8}_{-3.8}}

\newcommand{\LUMIN}{9.0~\mathrm{fb}^{-1}}

\newcommand{\LUMOUTL}{0.7~\mathrm{fb}^{-1}}
\newcommand{\LUMOUTH}{2.8~\mathrm{fb}^{-1}}

\begin{document}

\title{Observation of the decay \boldmath{$\x \to \pi^0 \chi_{c1}(1P)$}}

%\author{ W.~Imoehl$^{22}$, R.~E.~Mitchell$^{22}$, BESIII
%}

\author{
M.~Ablikim$^{1}$, M.~N.~Achasov$^{10,d}$, S. ~Ahmed$^{15}$, M.~Albrecht$^{4}$, M.~Alekseev$^{55A,55C}$,
A.~Amoroso$^{55A,55C}$, F.~F.~An$^{1}$, Q.~An$^{52,42}$, Y.~Bai$^{41}$, O.~Bakina$^{27}$
% ...
}

%\affiliation{
%~Institute of High Energy Physics, Beijing 100049, People's Republic of China\\
%$^{22}$ Indiana University, Bloomington, Indiana 47405, USA\\
%}

\affiliation{
Institute of High Energy Physics, Beijing 100049, People's Republic of China\\
% ...
$^{4}$ Bochum Ruhr-University, D-44780 Bochum, Germany\\
% ...
$^{10}$ G.I. Budker Institute of Nuclear Physics SB RAS (BINP), Novosibirsk 630090, Russia\\
% ...
$^{15}$ Helmholtz Institute Mainz, Johann-Joachim-Becher-Weg 45, D-55099 Mainz, Germany\\
% ...
$^{27}$ Joint Institute for Nuclear Research, 141980 Dubna, Moscow region, Russia\\
% ...
$^{41}$ Southeast University, Nanjing 211100, People's Republic of China\\
$^{42}$ State Key Laboratory of Particle Detection and Electronics, Beijing 100049, Hefei 230026, People's Republic of China\\
% ...
$^{52}$ University of Science and Technology of China, Hefei 230026, People's Republic of China\\
% ...
$^{55}$ (A)University of Turin, I-10125, Turin, Italy; (B)University of Eastern Piedmont, I-15121, Alessandria, Italy; (C)INFN, I-10125, Turin, Italy\\
% ...
$^{d}$ Also at the Novosibirsk State University, Novosibirsk, 630090, Russia\\
% ...
}

\collaboration{BESIII Collaboration}
\noaffiliation

\begin{abstract}
Using a total of $\LUMIN$ of $e^+e^-$ collision data with center-of-mass energies between 4.15 and 4.30~GeV collected by the BESIII detector, we search for the processes $e^+e^-\to \gamma \x$ with $\x\to\pi^0\chi_{cJ}$ for $J=0,1,2$.  
We report the first observation of $\x\to\pi^{0}\chi_{c1}$, a new decay mode of the $\x$, with a statistical significance of more than 5$\sigma$.
%$\FINALBSIGONE\sigma$.
Normalizing to the previously established process $e^+e^-\to \gamma \x$ with $\x\to\pi^+\pi^-J/\psi$, we find ${\cal B}(\x \to \pi^0 \chi_{c1})/{\cal B}(\x \to \pi^+\pi^- J/\psi) = \FINALBONE$, where the first error is statistical and the second is systematic.  
We set 90\% confidence level upper limits on the corresponding ratios for the decays to $\pi^0\chi_{c0}$ and $\pi^0\chi_{c2}$ of
$\FINALBULZERO$ 
and 
$\FINALBULTWO$, 
respectively.
\end{abstract}


\date{\today}
%\preprint{v16}

% LINE NUMBERS
%\begin{linenumbers}

\maketitle

In the mass region above open-charm threshold, where charmonium states are
heavy enough to decay to open-charm mesons, there are a number of
states with features that are yet to be satisfactorily
understood~\cite{reviews}.  These features likely point towards the
existence of non-$c\bar{c}$ configurations of charmonium.  The
$\x$~(also known as the $\chi_{c1}(3872)$) was the first of these
unexpected states to be discovered.  It was first observed in 2003 by
the Belle Collaboration in the process $B\to K\x$ with
$\x\to\pi^+\pi^-J/\psi$~\cite{xbelle}.  It has since been seen by many
other experiments in other processes and decay modes~\cite{pdg}.  Its
prominent features now include: its width is narrow
($\Gamma<1.2$~$\mevcc$)~\cite{xnarrow}; its mass is consistent with
the $D^0\bar{D}^{*0}$ threshold (with an error on the mass difference of
0.18~$\mevcc$)~\cite{pdg}; it has quantum numbers
$J^{PC}=1^{++}$~\cite{xjpc}; 
no isospin partners are currently known~\cite{xisospin}; 
it has isospin-violating decays since it
decays to both $\rho J/\psi$~\cite{xnarrow} and $\omega
J/\psi$~\cite{xomegajpsi}; it also decays to
$D^0\bar{D}^{*0}$~\cite{xdd}, $\gamma J/\psi$~\cite{xrad}, and $\gamma
\psi(2S)$~\cite{xrad}.  Despite this growing list of experimental
facts, the nature of the $\x$ remains unclear~\cite{reviews}.
Measuring pionic transitions of the $\x$ to the $\chi_{cJ}$ has been
proposed to be one way to distinguish between various interpretations.
If the $\x$ were a conventional $c\bar{c}$ state, transitions to the
$\chi_{cJ}$ should be very small (Ref.~\cite{voloshin2008} predicts $\Gamma(\x\to\pi^0\chi_{c1})\sim 0.06$~keV); 
if the $\x$ were a tetraquark or molecular state, on the other hand, these rates are expected to be sizeable~\cite{voloshin2008,mehen}.

The BESIII experiment, operating at the Beijing Electron Positron Collider~(BEPCII), previously observed the process $e^+e^-\to \gamma \x$ with $\x\to \pi^+\pi^-J/\psi$ using data collected at four center-of-mass energies ($\ecm$): 4.01, 4.23, 4.26, and 4.36~GeV~\cite{previous}.  
The cross section was shown to be largest at 4.23 and 4.26~GeV.  
Since that time, BESIII has collected more data in this energy region, including approximately 3~fb$^{-1}$ at 4.18~GeV and 0.5~fb$^{-1}$ at each of seven additional points between 4.19 and 4.27~GeV.  These additional data sets provide an opportunity to search for new decay modes of the $\x$ using the same production process $e^+e^- \to \gamma \x$.  
Data collected at different $\ecm$ can be combined and new $\x$ decays can be normalized to $e^+e^- \to \gamma \x$ with $\x\to \pi^+\pi^- J/\psi$, thereby canceling the production cross section and many systematic uncertainties.  

In this Letter, we report the first observation of the decay $\x\to\pi^0\chi_{c1}$ with a statistical significance of $\FINALBSIGONE\sigma$.  
Like the $\rho J/\psi$ decay, this final state has an isospin of one.  
This is the first observation of a decay of the $\x$ to a $P$-wave charmonium state 
and its large branching fraction relative to $\pi^+\pi^- J/\psi$ supports
a non-$c\bar{c}$ interpretation of the $\x$~\cite{voloshin2008,mehen}.

We search for the processes $e^+e^-\to \gamma_1 \x$ with
$\x\to\pi^0\chi_{cJ}$~($J=0,1,2$) using the decays
$\chi_{cJ}\to\gamma_2 J/\psi$ and $J/\psi \to l^+l^-$, where $l^+l^-$
denotes both $e^+e^-$ and $\mu^+\mu^-$, and $\gamma_{1}$ and
$\gamma_{2}$ are the initial photon and the photon from the
$\chi_{cJ}$ decay, respectively.  This is subsequently referred to as
the ``search'' channel and it results in the final state
$\gamma_1\gamma_2\pi^0 l^+l^-$~(with $\pi^0\to\gamma\gamma$).  We also
reconstruct the ``normalization'' channel $e^+e^-\to \gamma \x$ with
$\x\to\pi^+\pi^-J/\psi$ and $J/\psi \to l^+l^-$, resulting in the
final state $\gamma\pi^+\pi^-l^+l^-$.  For the signal region, we use
all available BESIII data with $\ecm$ between 4.15 and
4.30~GeV~($\LUMIN$), where the $e^+e^-\to \gamma \x$ cross section was
measured to be largest; and for the sideband regions, we use all data
with $\ecm$ between 4.00 and 4.15~GeV~($\LUMOUTL$) and between 4.30
and 4.60~GeV~($\LUMOUTH$).

The Beijing Spectrometer~(BESIII) experiment uses a general purpose magnetic spectrometer \cite{BESIIIHardware}.  A superconducting solenoid magnet provides a 1.0~T magnetic field. Enclosed within the magnet are a helium-gas-based drift chamber~(MDC) for charged particle tracking
and a CsI(Tl) Electromagnetic Calorimeter~(EMC) to measure the energy of electromagnetic showers. Other detector components, such as the plastic scintillator time-of-flight system~(TOF), are not used in this analysis.

A \textsc{geant4}-based~\cite{Geant4} Monte Carlo~(MC) simulation
package is used to determine detection efficiencies and estimate
background rates.  The initial $e^+e^-$ collisions, including effects
due to Initial State Radiation~(ISR), and subsequent decays are
simulated using \textsc{kkmc}~\cite{KKMC} and
\textsc{evtgen}~\cite{EvtGen}, respectively.
Final State Radiation~(FSR) is simulated with \textsc{PHOTOS}~\cite{PHOTOS}.

Optimization of the event selection criteria is performed using three categories of data samples: one to estimate signal yields~($S$), and two for background yields~($B_1$ and $B_2$).  For $S$, signal MC samples are used. The normalization channel is generated so that  
$\sigma(e^+e^-\to\gamma\x)\times{\cal B}(\x \to \pi^+\pi^- J/\psi) = 0.3$~pb
at each $\ecm$~\cite{previous};
the search channels are initially scaled assuming ${\cal B}(\x \to \pi^0 \chi_{cJ})/{\cal B}(\x \to \pi^+\pi^- J/\psi) = 1$.
For $B_1$, background MC samples for processes including a $J/\psi$ are generated using previously measured cross sections.  These include 
$e^+e^-\to\pi\pi J/\psi$~\cite{pipijpsi,pizpizjpsi}, 
$\pi\pi\psi(3686)$~\cite{pipipsip,pizpizpsip}, 
$\eta J/\psi$~\cite{etajpsi}, 
$\eta^\prime J/\psi$~\cite{etapjpsi}, 
$\omega \chi_{cJ}$~\cite{omegachic,omegachicb}, and 
$\gamma_{\mathrm{ISR}}\psi(3686)$~\cite{isr,psipeewidth}.  
These also include $e^+e^-\to \gamma \x$ with $\x$ decays to 
$\omega J/\psi$~\cite{xomegajpsi}, 
$\gamma J/\psi$~\cite{xrad} and 
$\gamma \psi(3686)$~\cite{xrad}, each of which is normalized to $\x\to\pi^+\pi^-J/\psi$ using previous measurements~\cite{pdg}.  
For $B_2$, background modes that do not include a $J/\psi$ are estimated using sidebands in the reconstructed mass spectrum of $J/\psi$ candidates in data.  Analysis of an inclusive MC sample shows no other background modes with peaks near the $J/\psi$, $\chi_{cJ}$, or $\x$ masses.

Common charged particle and photon selection criteria are used for the normalization and search channels.  Charged particles are selected using their distance of closest approach to the interaction region (within 10~cm along the beam direction and 1~cm transverse to it) and are required to be within the region  $|\cos\theta|< 0.93$, where $\theta$ is measured with respect to the beam axis.  No particle identification is used for charged pions.  Electrons and muons are distinguished using the energy they deposit in the EMC divided by their momentum ($E/p$): 
charged tracks are labeled as electrons~(muons) in the case $E/p>0.85$~($E/p<0.25$), respectively.
Photons must have deposited an energy greater than 25~MeV in the barrel region of the EMC~($|\cos\theta|< 0.80$) and greater than 50~MeV in the endcap region~($0.86<|\cos\theta|< 0.92$), and must have a hit time within 700~ns of the event start time.

Using the selected charged particles and photons, kinematic fits are
then performed for the normalization
channel~($\gamma\pi^+\pi^-l^+l^-$) and search
channel~($\gamma_1\gamma_2\pi^0l^+l^-$) hypotheses.  A
four-constraint~(4C) kinematic fit is used for the normalization
channel, where the total measured four-momentum is constrained to the
four-momentum of the
initial center-of-mass system, and the resulting $\chidoffour$ is
required to be less than 10.  For the search channel, an extra
constraint~(1C) is added to constrain a $\gamma\gamma$ pair to the
$\pi^0$ mass and we require $\chidoffive<5$.  These criteria are optimized
by maximizing $S/\sqrt{S+B_1+B_2}$, where the sizes of the
signal~($S$) and background~($B_1$ and $B_2$) are determined from the
three data samples described previously.  Multiple combinations per
event are allowed, but are negligible after event selection.  Using
signal MC samples, multiply counted events are found to be less than
0.1\% and 4\% in the normalization and search channels, respectively.
In data, no multiply counted events are found.

\begin{figure}[t]
\centerline{\includegraphics[clip=true, viewport=0 55 565 420, width=1.0\columnwidth]{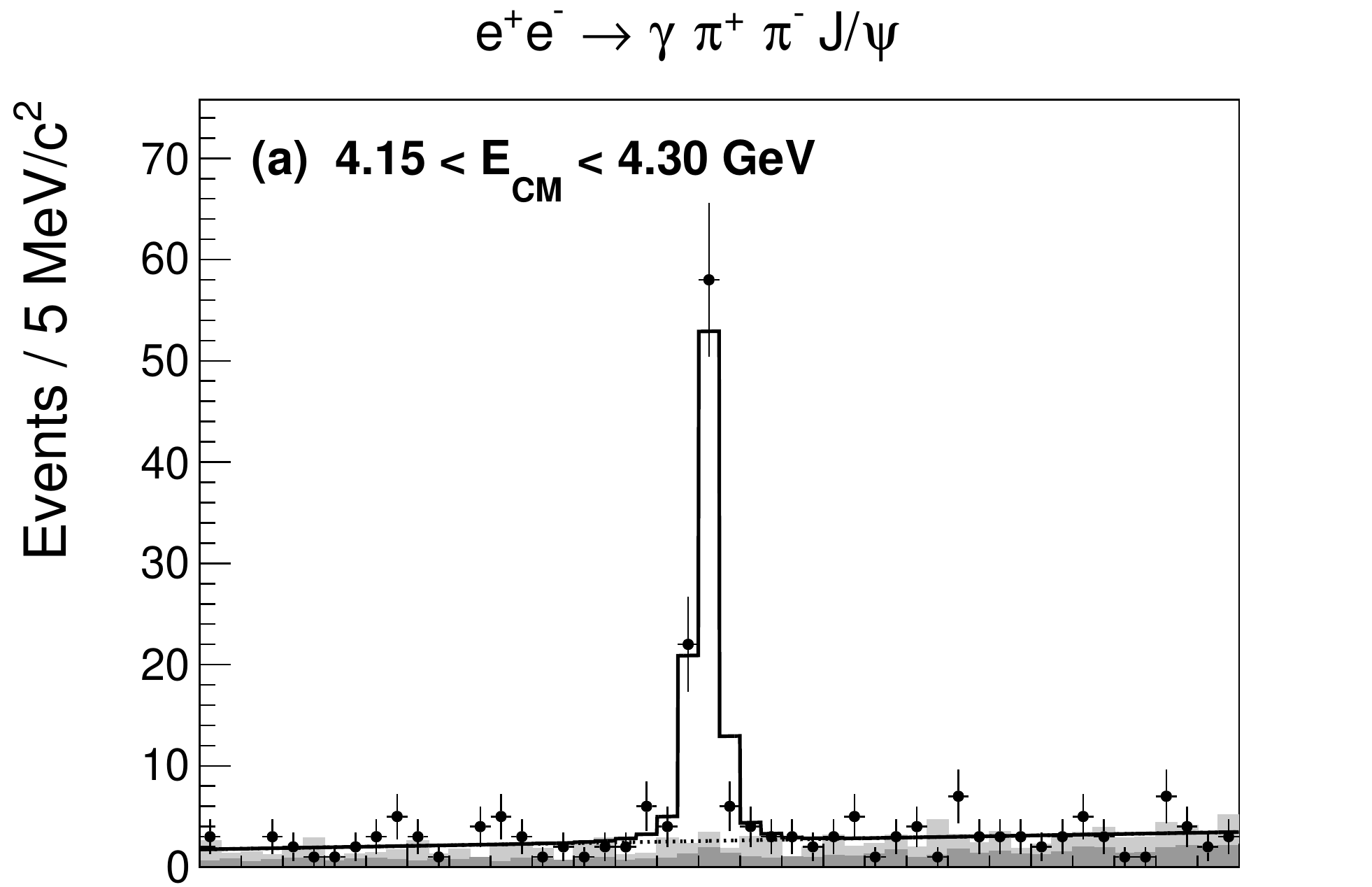}}
\centerline{\includegraphics[clip=true, viewport=0 0 565 385, width=1.0\columnwidth]{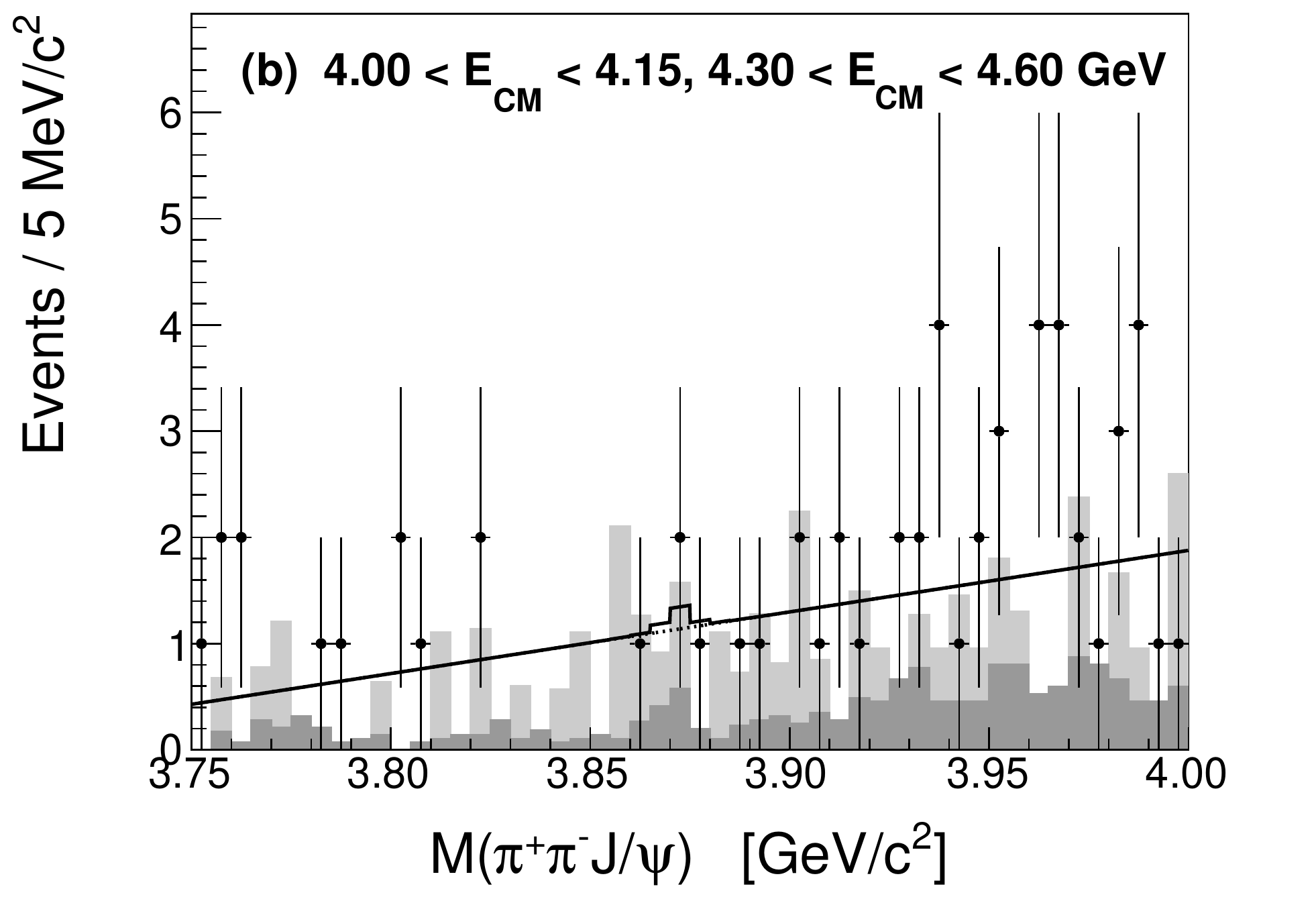}}
\caption{\label{fig:norm}
Distribution of $\pi^+\pi^-J/\psi$ mass, $M(\pi^+\pi^-J/\psi)$, from the normalization process $e^+e^-\to\gamma\pi^+\pi^-J/\psi$ for (a)~$4.15 < \ecm < 4.30$~GeV and (b)~$4.00 < \ecm < 4.15$ or $4.30 < \ecm < 4.60$~GeV.  Points are data; lines are fits (solid is the total and dotted is the polynomial background); the darker histogram is a MC estimate of peaking $J/\psi$ backgrounds; the lighter stacked histogram is an estimate of non-peaking backgrounds using $J/\psi$ sidebands from data.}
\end{figure}

\begin{figure}[t]
\centerline{\includegraphics[clip=true, viewport=0 55 565 420, width=1.0\columnwidth]{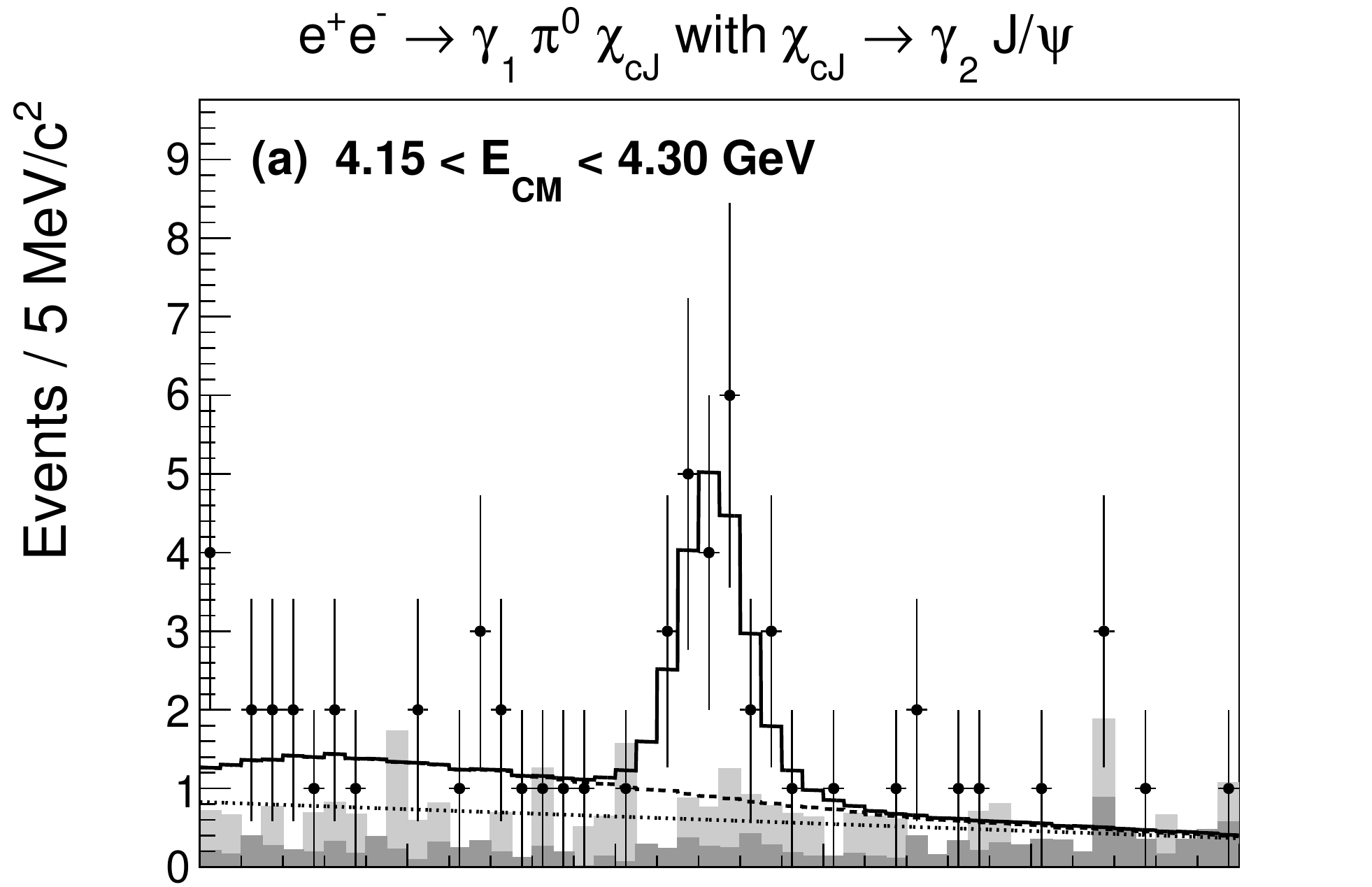}} 
\centerline{\includegraphics[clip=true, viewport=0 0 565 385, width=1.0\columnwidth]{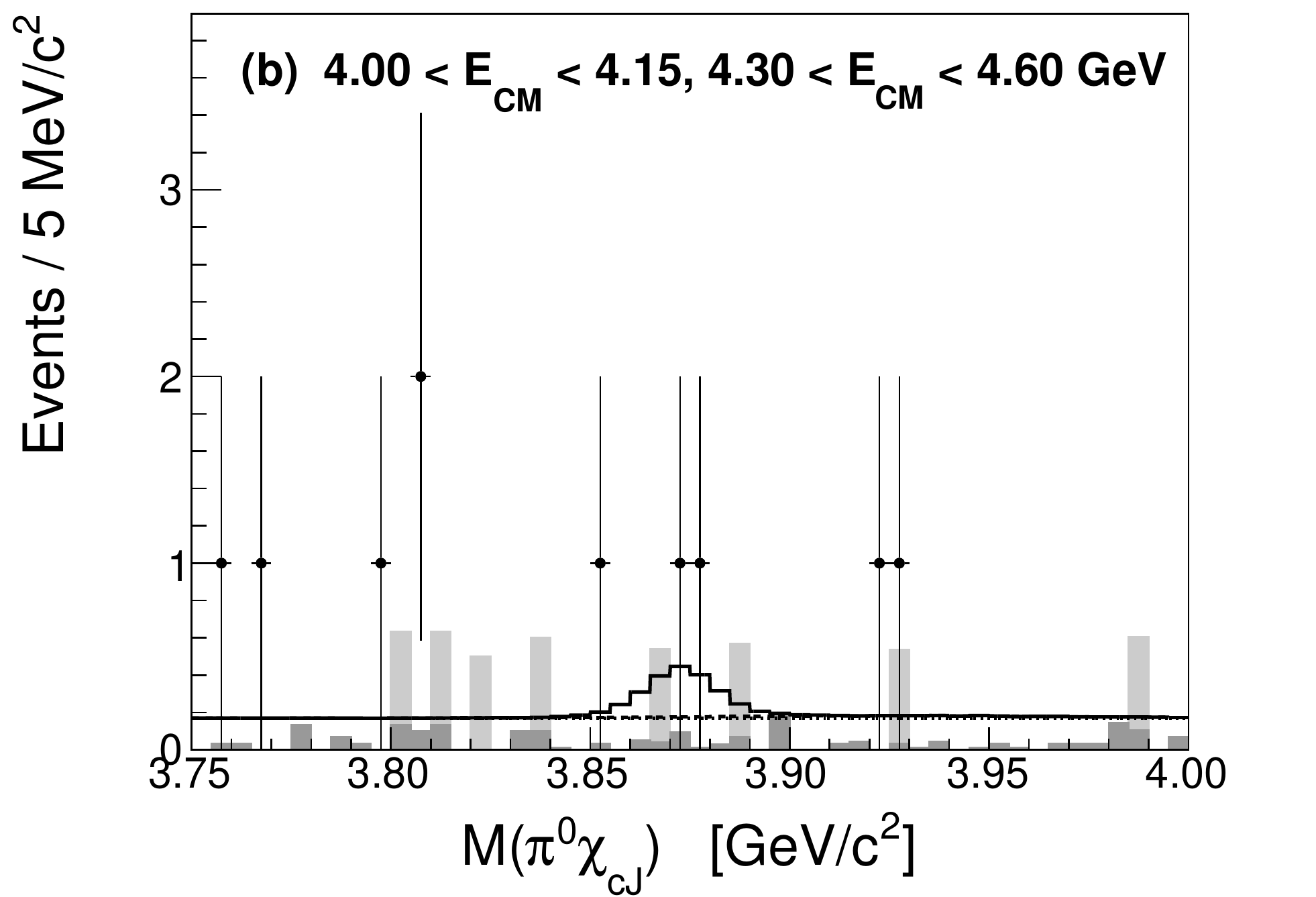}} 
\caption{\label{fig:search}
Distribution of $\pi^0\chi_{cJ}$ mass, $M(\pi^0\chi_{cJ})$, from the process $e^+e^-\to\gamma\pi^0\chi_{cJ}$ for (a)~$4.15 < \ecm < 4.30$~GeV and (b)~$4.00 < \ecm < 4.15$ or $4.30 < \ecm < 4.60$~GeV. The $\chi_{cJ}$ are selected using a broad region of $\gamma J/\psi$ mass. Points, lines, and histograms follow the same convention as Fig.~\ref{fig:norm}. The dashed line is the total background contribution to the fit, including signal events with $\gamma_1$ and $\gamma_2$ interchanged.}
\end{figure}

The $J/\psi$ signal is selected by requiring $M(l^+l^-)$ to be within 20~$\mevcc$ of the nominal $J/\psi$ mass~\cite{pdg}.  The $J/\psi$ sideband regions, used for background estimations, are each 40~$\mevcc$ wide on either side of the $J/\psi$ and leave a 20~$\mevcc$ gap with the signal region.

Several additional criteria are used to select the normalization channel.  
Radiative Bhabha background events ($e^+e^-\to e^+e^-(n\gamma)$), where a radiated photon converts to $e^+e^-$ within the detector material and the resulting $e^+e^-$ are 
mistaken to be
$\pi^+\pi^-$, are removed by requiring the $\pi^+\pi^-$ opening angle ($\theta_{\pi\pi}$) to satisfy $\cos\theta_{\pi\pi}< 0.98$.  Further suppression of this background process is obtained by requiring the opening angle of the final-state photon and any charged track~($\theta_{\gamma\mathrm{tk}}$) 
to satisfy $\cos\theta_{\gamma\mathrm{tk}}< 0.98$.  Background events from $\eta J/\psi$ and $\eta^\prime J/\psi$ are removed by requiring $M(\gamma\pi^+\pi^-)>0.6~\gevcc$ and
$|M(\gamma\pi^+\pi^-)-M_0(\eta^\prime)|>0.02~\gevcc$ ($M_0(\eta^\prime)$ is the nominal mass of the $\eta^\prime$~\cite{pdg}), respectively.

For the search channel, the background mode $\pi^0\pi^0J/\psi$ is suppressed both by 
requiring  $M(\gamma_1\gamma_2)$ to be 20~$\mevcc$ away from the $\pi^0$ mass and by placing the same requirement on the mass of $\gamma_1$ or $\gamma_2$ combined with the higher energy photon from the $\pi^0$ decay.  Background events from $\omega(782)$ decays to $\gamma\pi^0$, including those from $e^+e^-\to\omega\chi_{cJ}$ and $\gamma\x\to\gamma\omega J/\psi$, are removed by requiring $M(\gamma_{1,2}\pi^0)<0.732~\gevcc$.
Finally, background events from $\gamma_{\mathrm{ISR}}\psi(3686)$ are reduced by requiring the mass recoiling against $\gamma_1$ or $\gamma_2$ both to be larger than 3.7~$\gevcc$.

The final distributions for the reconstructed $\pi^+\pi^-J/\psi$ mass
in the normalization channel are shown in Fig.~\ref{fig:norm}.  
In order to improve the mass resolution, $M(\pi^+\pi^-J/\psi)$ is
calculated using $M(\pi^+\pi^-l^+l^-) - M(l^+l^-) + M_0(J/\psi)$, 
where $M_0(J/\psi)$ is the nominal mass of the $J/\psi$.
The mass resolution is improved from 7.4~MeV/$c^2$ to 4.7~MeV/$c^2$.
Figure~\ref{fig:norm}a corresponds to data taken
at $4.15<\ecm<4.30$~GeV and shows a clear $\x$ signal.  The data are
fitted by a first-order polynomial representing the background and a
response function of the signal process that has been obtained from
the signal MC simulation.  All fits are performed using a binned
likelihood method; all significances are obtained by comparing the
resulting likelihoods with and without the signal component included.
Results are listed in Table~\ref{tab:results}.  Figure~\ref{fig:norm}b
shows the same for the other $\ecm$ samples.  No $\x$ signal is seen.
This pattern is consistent with the previous
measurement~\cite{previous}.

\begin{figure}[t!]
\centerline{\includegraphics[clip=true, viewport=0 0 565 420, width=1.0\columnwidth]{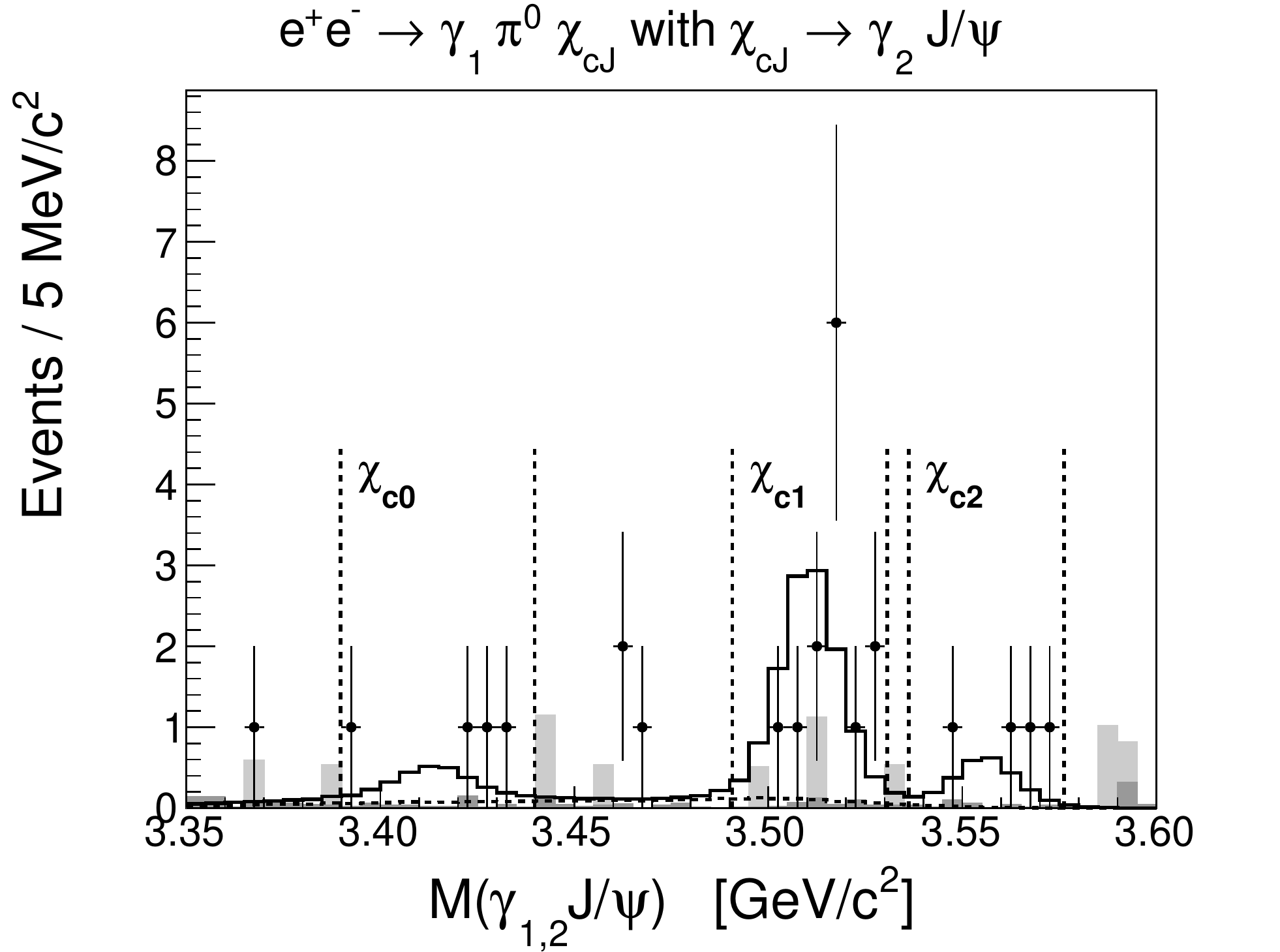}} 
\caption{\label{fig:chicj}
Distribution of $M(\gamma_{1,2}J/\psi)$ after selecting the $\x$ signal region from Fig.~\ref{fig:search}a.
Points and shaded histograms follow the same convention as Fig.~\ref{fig:norm}. The solid line is the signal~MC and is scaled using subsequent fits; the dashed line is the component of the signal~MC where
$\gamma_1$ and $\gamma_2$ are interchanged.  Vertical lines show the $\chi_{cJ}$ selection regions.}
\end{figure}

The corresponding distributions of $M(\pi^0\chi_{cJ})$ for the search
channel are shown in Fig.~\ref{fig:search}.  The $\chi_{cJ}$ region is
first chosen with a loose requirement on $M(\gamma_{1,2}J/\psi) \equiv
M(\gamma_{1,2}l^+l^-)- M(l^+l^-) + M_0(J/\psi)$ between 3.35 and
3.60~$\gevcc$.  A clear signal for the $\x$ is observed for
$4.15<\ecm<4.30$~GeV~(Fig.~\ref{fig:search}a); no evidence for the
$\x$ is seen at other $\ecm$~(Fig.~\ref{fig:search}b).  The
distributions are fit with a first-order polynomial background function
and a signal shape derived from the signal MC simulation, where the
relative fractions of $\pi^0\chi_{cJ}$ with $J=0,1,2$ are fixed by
subsequent fits.  There are two entries per event corresponding to the
two combinations of $\gamma_1$ and $\gamma_2$; the signal MC includes
a broad contribution from events with interchanged $\gamma_1$ and
$\gamma_2$.  Using the background samples described earlier~($B_1$ and
$B_2$), we find no other peaking background events.  The fit in
Fig.~\ref{fig:search}a yields $\FINALNCHICJ$ $X(3872)$ events with a
statistical significance of $\FINALBSIGCHICJ\sigma$.

\begin{figure}[t!]
\centerline{\includegraphics[clip=true, viewport=0 55 565 420, width=1.0\columnwidth]{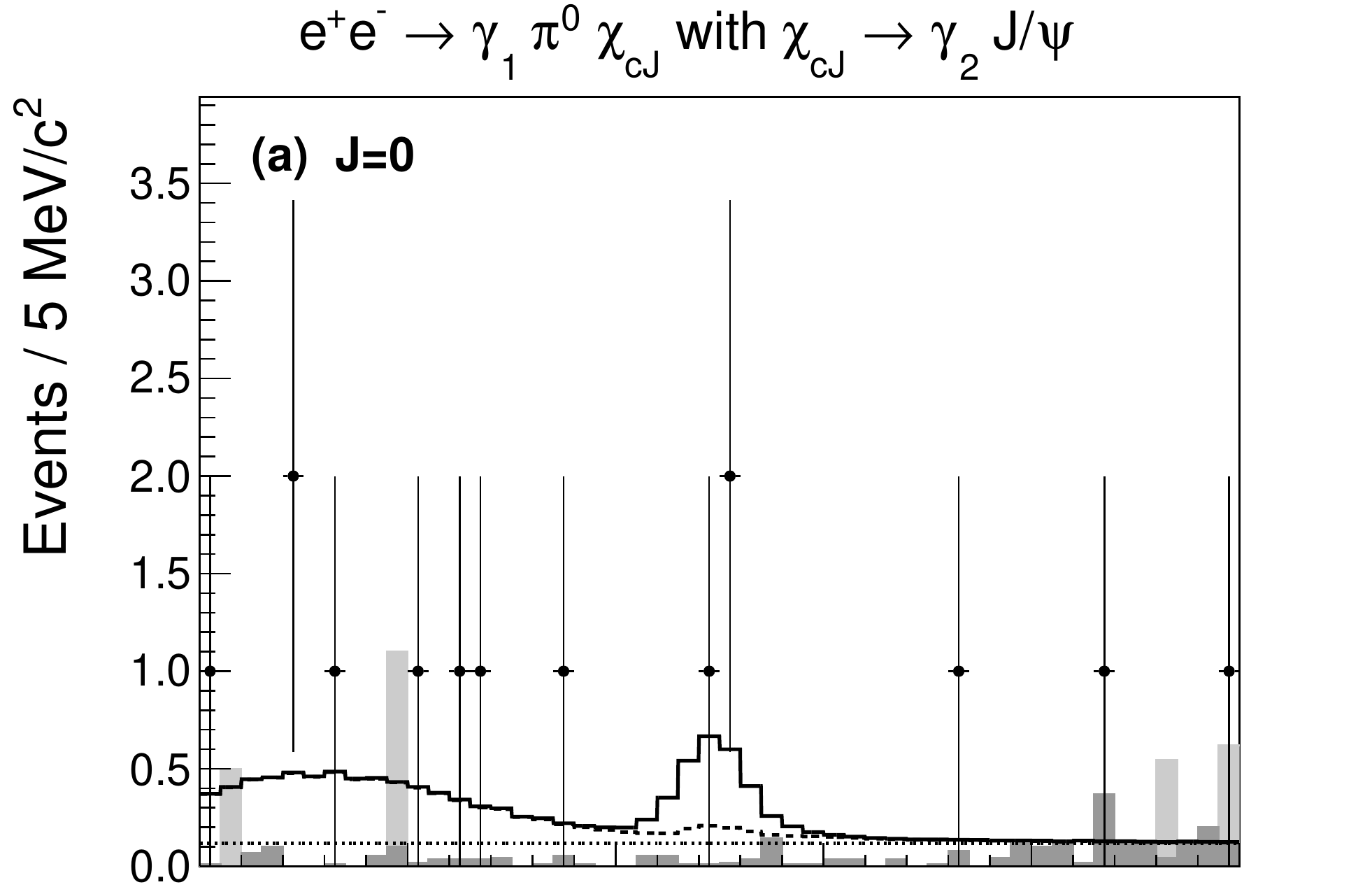}} 
\centerline{\includegraphics[clip=true, viewport=0 55 565 385, width=1.0\columnwidth]{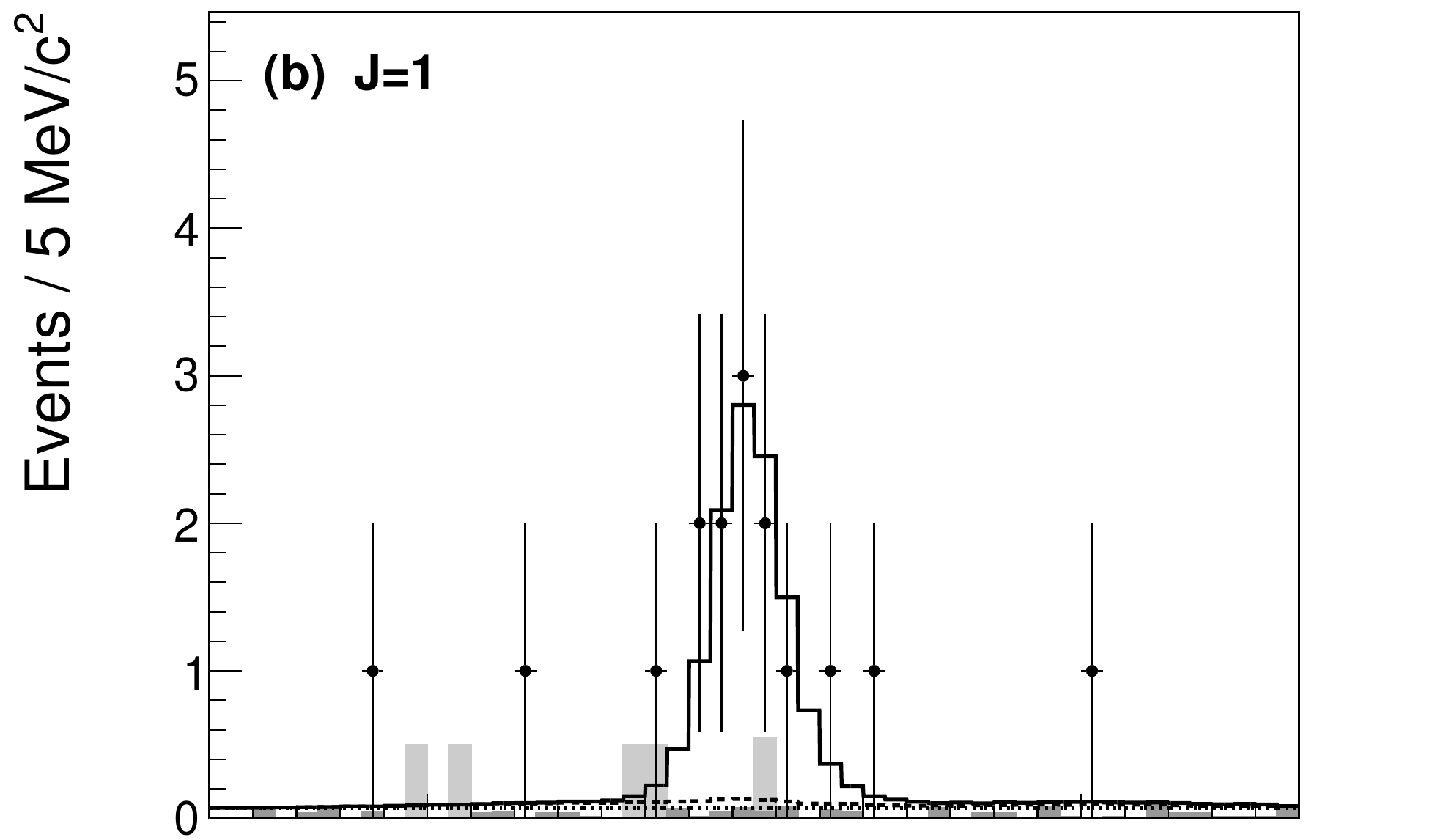}} 
\centerline{\includegraphics[clip=true, viewport=0 0 565 385, width=1.0\columnwidth]{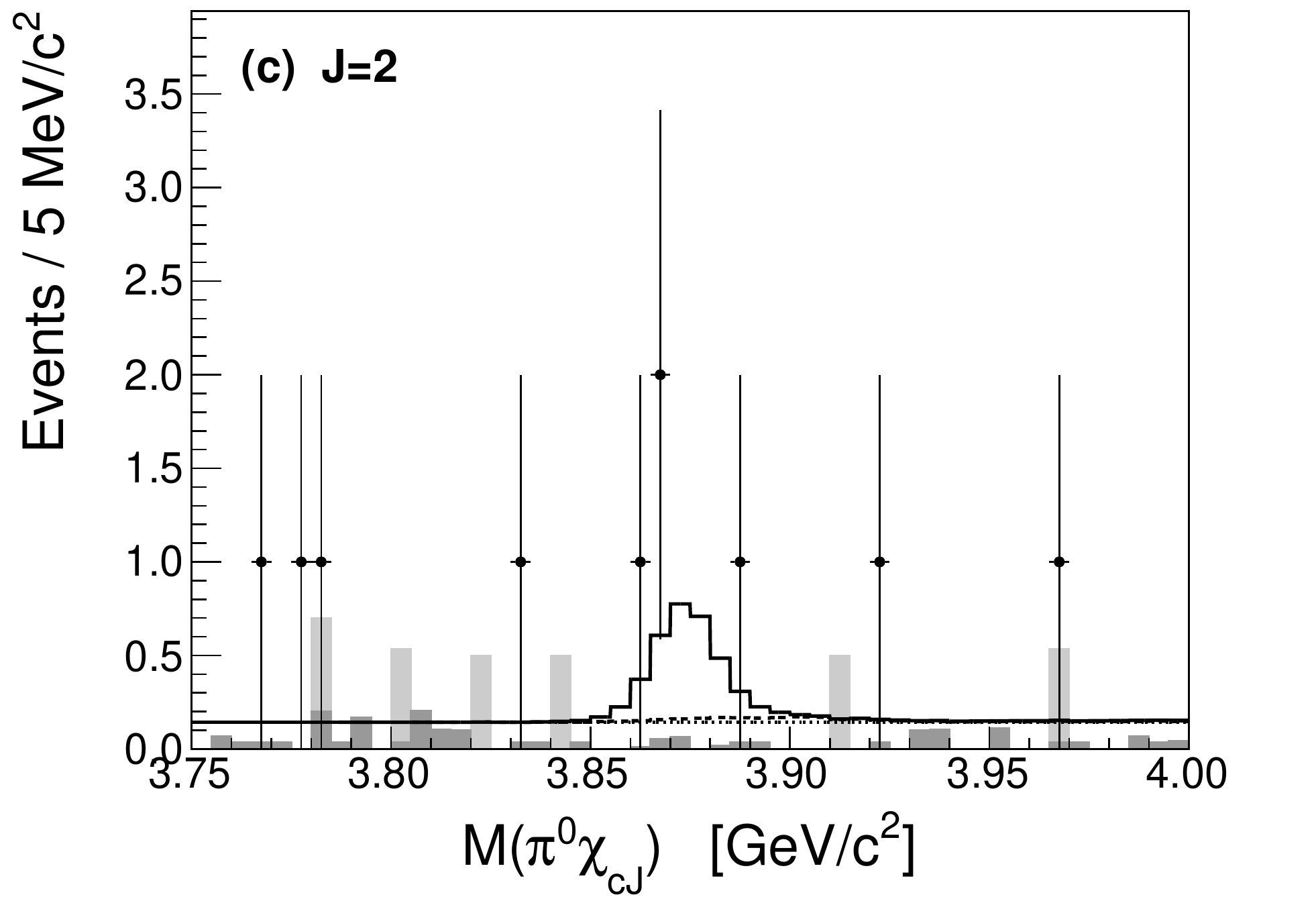}}
\caption{\label{fig:fit}
Distributions of $\pi^0\chi_{cJ}$ mass, $M(\pi^0\chi_{cJ})$, from the process $e^+e^-\to\gamma\pi^0\chi_{cJ}$ for (a)~$J=0$, (b)~$J=1$, and (c)~$J=2$. Points, lines, and histograms follow the same convention as Fig.~\ref{fig:norm}.  The dashed line is the total background in the fit and includes contributions from events with interchanged $\gamma_1$ and $\gamma_2$ and cross-feed among the search channels. 
}
\end{figure}

We next use the $M(\gamma_{1,2}J/\psi)$ distribution to select the
$\chi_{c0}$, $\chi_{c1}$, and $\chi_{c2}$ mass
regions~(Fig.~\ref{fig:chicj}).  The photons $\gamma_1$ and $\gamma_2$
are separated by choosing $\gamma_2$ to be the photon that minimizes
$\Delta M_J \equiv |M(\gamma_2 J/\psi)-M_0(\chi_{cJ})|$, where
$M_0(\chi_{cJ})$ is the nominal mass of each $\chi_{cJ}$~\cite{pdg}.
We require $\Delta M_0 < 25~\mevcc$ and $\Delta M_{1,2} < 20~\mevcc$.
The resulting distributions for $M(\pi^0\chi_{cJ})$ with $J=0,1,2$ are
shown in Fig.~\ref{fig:fit}.  Each $M(\pi^0\chi_{cJ})$ distribution is
fit with a constant background function and a signal shape derived
from signal~MC simulation.  The signal MC samples include events with
interchanged $\gamma_1$ and $\gamma_2$ as well as cross-feed among the
$\pi^0\chi_{cJ}$ channels.  These effects result in an additional peak
below the $\x$ signal region in the $M(\pi^0\chi_{c0})$ distribution,
but are negligible elsewhere.  In the $M(\pi^0\chi_{c1})$
distribution, we find a $\x$ signal with a $\FINALBSIGONE\sigma$
significance.  No significant $\x$ signals are found in the
$M(\pi^0\chi_{c0,2})$ distributions.  Numbers for events,
efficiencies, and significances are listed in Table~\ref{tab:results}.
The total yield of signal events in all three channels is $\TOTALNJ$,
consistent with the fit in Fig.~\ref{fig:search}a.

Also shown in Table~\ref{tab:results} are the final ratios ${\cal B}(\x\to\pi^{0}\chi_{cJ})/{\cal B}(\x\to\pi^{+}\pi^{-}J/\psi)$.  These are calculated from the ratios of yields of signal events, the ratios of efficiencies~(including minor effects due to ISR), and the nominal $\chi_{cJ}$ and $\pi^0$ branching fractions~\cite{pdg}.  
Upper limits (at the 90\% C.L.) are calculated from the likelihood curve of the fits as a function of signal yield after being convolved with a Gaussian distribution with a width the size of the systematic uncertainty. 
The $J/\psi$ branching fractions, integrated luminosities at each $\ecm$, ISR correction factors, as well as a number of systematic uncertainties cancel in the ratios.

\begin{table*}[t]
\caption{\label{tab:results}
Final results for the normalization and search channels and their ratios.  Individual efficiencies are reported without considering ISR in the MC~(no ISR) and are for illustration only. Efficiency ratios are for the search channels divided by the normalization channel and include effects due to ISR~(with ISR), which nearly cancel in the ratio.  Numbers in parentheses are 90\% C.L. upper limits.  The first errors are statistical and the second are systematic.
}
\begin{ruledtabular}
\begin{tabular}{lcccc}
  &  $\pi^{+}\pi^{-}J/\psi$  &  $\pi^{0}\chi_{c0}$  &  $\pi^{0}\chi_{c1}$  &  $\pi^{0}\chi_{c2}$  \\
\hline
\noalign{\vskip 1mm}
Event yield  &  $84.1^{+10.1}_{-9.4}$  &  $1.9^{+1.9}_{-1.3}$  &  $10.8^{+3.8}_{-3.1}$  &  $2.5^{+2.3}_{-1.7}$  \\
Signal significance ($\sigma$)  &  16.1  &  1.6  &  5.2  &  1.6  \\
Efficiency (no ISR) (\%)  &  32.3  &  8.8  &  14.1  &  12.8  \\
Efficiency ratio (with ISR)  & ... &  0.272  &  0.435  &  0.392  \\
${\cal B}(\chi_{cJ}\to\gamma J/\psi)\times{\cal B}(\pi^{0}\to\gamma\gamma)$ (\%)  &  ...  &  1.3  &  33.5  &  19.0  \\
Total systematic error (\%)  & ... &  17.0  &  11.9  &  9.4  \\
${\cal B}(X\to\pi^{0}\chi_{cJ})/{\cal B}(X\to\pi^{+}\pi^{-}J/\psi)$  & ... &  $6.6^{+6.5}_{-4.5}\pm1.1$ (19)  &  $0.88^{+0.33}_{-0.27}\pm0.10$  &  $0.40^{+0.37}_{-0.27}\pm0.04$ (1.1)  \\
\end{tabular}

\end{ruledtabular}
\end{table*}

\begin{table}[ht]
\caption{\label{tab:syst} 
Relative systematic uncertainties on the ratio ${\cal B}(\x \to \pi^0 \chi_{cJ})/{\cal B}(\x \to \pi^+\pi^- J/\psi)$ for $J = 0, 1, 2$.  All entries are in percent.
}
\begin{ruledtabular}
\begin{tabular}{lccc}
  &  $\pi^{0}\chi_{c0}$  &  $\pi^{0}\chi_{c1}$  &  $\pi^{0}\chi_{c2}$  \\
\hline
(1)~Photon efficiencies  &  3.0  &  3.0  &  3.0  \\
(2)~Track efficiencies  &  2.0  &  2.0  &  2.0  \\
(3)~Input branching fractions  &  4.7  &  3.5  &  3.6  \\
(4)~Kinematic fit  &  4.6  &  4.6  &  4.6  \\
(5)~$\ecm$-dependence of efficiency ratio  &  3.2  &  5.2  &  5.2  \\
(6)~MC decay models  &  8.2  &  8.1  &  2.3  \\
(7)~Fitting to determine signal yield &  12.4  &  1.6  &  3.0  \\
\hline
Total  &  17.0  &  11.9  &  9.4  \\
\end{tabular}
 
\end{ruledtabular}
\end{table}

The remaining systematic uncertainties are listed in
Table~\ref{tab:syst}.  (1,2)~For uncertainties in the photon and
charged track efficiencies, we use 1\% per
photon~\cite{Ablikim:2010zn} and track~\cite{etapjpsi} that do not
cancel between the search and normalization channels.  (3)~For input
branching fractions, uncertainties from the PDG are used~\cite{pdg}.
(4)~A systematic uncertainty due to the kinematic fit is
determined using clean control samples with matching final states:
$e^+e^-\to\pi^0\pi^0J/\psi$ for the search channel and
$e^+e^-\to\gamma_{\mathrm{ISR}}\psi(2S)\to\gamma_{\mathrm{ISR}}\pi^+\pi^-J/\psi$
for the normalization channel.  (5)~The selection criteria that
distinguish between $\gamma_1$ and $\gamma_2$ in the search channel
introduce some $\ecm$-dependence in the efficiency ratio.  To probe
this uncertainty, we generate different shapes for the cross section
as a function of $\ecm$: the nominal is constant, one is based on the
$e^+e^-\to\pi^+\pi^-J/\psi$ lineshape seen by BESIII~\cite{pipijpsi},
and one is based on the $\psi(4160)$ lineshape with parameters from the
PDG~\cite{pdg}.  We take the largest difference as a systematic
uncertainty.  (6)~Signal~MC samples are generated according to
realistic spin-dependent amplitudes using
\textsc{evtgen}~\cite{EvtGen}. In channels where there is ambiguity
(e.g. the presence of both $S$- and $D$-waves in $\x\to\rho J/\psi$~\cite{xnarrow} or
both $P$- and $F$-waves in $\x\to\pi^0\chi_{c2}$), we replace our
nominal models by phase space and take the maximum difference as a
systematic uncertainty.  (7)~Fitting uncertainties are evaluated using
two fit variations: zeroth- and first-order background polynomials, and
a signal shape that is widened by 20\% to account for possible
differences in mass resolution between data and MC~simulation.  The
significance of the signal for $\x\to\pi^0\chi_{c1}$ remains above
5$\sigma$ for all variations.  The total systematic uncertainty is
obtained by adding the individual uncertainties in quadrature.

In summary, we use $\LUMIN$ of $e^+e^-$ collision data with $\ecm$ between 4.15 and 4.30~GeV to search for the processes $e^+e^- \to \gamma \x$ with $\x \to \pi^0 \chi_{cJ}$.  
We make the first observation of the process $\x \to \pi^0 \chi_{c1}$, where the statistical significance is greater than $5\sigma$ for all systematic variations.
Normalizing to $e^+e^- \to \gamma \x$ with $\x \to \pi^+ \pi^- J/\psi$, we determine the ratio ${\cal B}(\x \to \pi^0 \chi_{c1})/{\cal B}(\x \to \pi^+\pi^- J/\psi) = \FINALBONE$.
Upper limits (at the 90\% C.L.) for the corresponding ratios for the $\pi^0\chi_{c0}$ and $\pi^0\chi_{c2}$ decays are $\FINALBULZERO$ 
and 
$\FINALBULTWO$, 
respectively.
Using ${\cal B}(\x \to \pi^+\pi^- J/\psi) > 3.2\%$~\cite{pdg}
~(obtained by comparing exclusive~\cite{exclusive} and inclusive~\cite{inclusive} $B^+$ decays)
and ${\cal B}(\x \to \pi^+\pi^- J/\psi) < 6.4\%$
~(obtained by assuming all measured $\x$ decays add to less than 100\%),
we find ${\cal B}(\x \to \pi^0 \chi_{c1}) \sim 3 - 6\%$.
If the $\x$ were the $\chi_{c1}(2P)$ state of charmonium, Ref.~\cite{voloshin2008} predicts $\Gamma(\x\to\pi^0\chi_{c1})\sim 0.06$~keV.
Combining this with our result, this would imply a total width of the $\x$ of only $\sim 1.0 - 2.0$~keV, which would be orders of magnitude smaller than all other observed charmonium states.
Therefore, our measurement disfavors the $c\bar{c}$ interpretation of the $\x$.

The BESIII collaboration thanks the staff of BEPCII and the IHEP computing center for their strong support. This work is 
supported in part by National Key Basic Research Program of China under Contract No. 2015CB856700; National Natural Science 
Foundation of China (NSFC) under Contracts Nos. 11335008, 11425524, 11625523, 11635010, 11735014; the Chinese Academy of 
Sciences (CAS) Large-Scale Scientific Facility Program; the CAS Center for Excellence in Particle Physics (CCEPP); Joint 
Large-Scale Scientific Facility Funds of the NSFC and CAS under Contracts Nos. U1532257, U1532258, U1732263; CAS Key 
Research Program of Frontier Sciences under Contracts Nos. QYZDJ-SSW-SLH003, QYZDJ-SSW-SLH040; 100 Talents Program of CAS; 
INPAC and Shanghai Key Laboratory for Particle Physics and Cosmology; German Research Foundation DFG under Contract No. 
Collaborative Research Center CRC 1044; Istituto Nazionale di Fisica Nucleare, Italy; Koninklijke Nederlandse Akademie van 
Wetenschappen (KNAW) under Contract No. 530-4CDP03; Ministry of Development of Turkey under Contract No. DPT2006K-120470; 
National Science and Technology fund; The Swedish Research Council; the Knut and Alice Wallenberg foundation; U. S. Department of Energy under Contracts Nos. 
DE-FG02-05ER41374, DE-SC-0010118, DE-SC-0012069; University of Groningen (RuG) and the Helmholtzzentrum fuer 
Schwerionenforschung GmbH (GSI), Darmstadt.

\end{document}